# Representing Geographic Space as a Hierarchy of Recursively Defined Subspaces for Computing the Degree of Order

Bin Jiang and Chris de Rijke

Faculty of Engineering and Sustainable Development, Division of GIScience
University of Gävle, SE-801 76 Gävle, Sweden
Email: bin.jiang|chris.de.rijke@hig.se

*What is needed is to develop a new mathematics of order and structure. This requires an extensive study, in which one slowly and carefully 'feels one's way' into the subject. It cannot properly be done solely by applying existing mathematics, because the latter does not have the right general structure.*
David Bohm (1969)

*The activity we call building creates the physical order of the world, constantly, unendingly, day after day ... Our world is dominated by the order we create.*
Christopher Alexander (2002–2005)

**Abstract**
As Christopher Alexander discovered, all space or matter – either organic or inorganic – has some degree of order in it according to its structure and arrangement. The order refers to a kind of structural character, called living structure, which is defined as a mathematical structure that consists of numerous substructures with an inherent hierarchy. Across the hierarchy, there are far more small substructures than large ones, while on each level of the hierarchy the substructures are more or less similar in size. In this paper we develop a new approach to representing geographic space as a hierarchy of recursively defined subspaces for computing the degree of order. A geographic space is first represented as a hierarchy of recursively defined subspaces, and all the subspaces are then topologically represented as a network for computing the degree of order of the geographic space, as well as that of its subspaces. Unlike conventional geographic representations, which are mechanical in nature, this new geographic representation is organic, conceived, and developed under the third view of space; that is, space is neither lifeless nor neutral, but a living structure capable of being more living or less living. Thus, the order can also be referred to as life, beauty, coherence, or harmony. We applied the new representation to three urban environments, 253 patterns, and 35 black-white strips to verify it and to demonstrate advantages of the new approach and the new kind of order. We further discuss the implications of the approach and the order on geographic information science and sustainable urban planning.

**Keywords:** Living structure, pattern language, life, wholeness, coherence, structural beauty

## 1. Introduction
In his life's work, *The Nature of Order*, Christopher Alexander (2002–2005) observed and proved that any space or matter has a certain degree of order in it according to its structure and arrangement. A space or matter with a high degree of the order looks alive or vibrant. In his earlier work, *The Timeless Way of Building* (Alexander 1979), Alexander referred to the kind of order as *"quality without a name"*, which is synonymous to the notion of vitality or organized complexity (Jacobs 1961). The *"quality without a name"* has some obscure characteristics that may be captured by synonyms such as alive, whole, comfortable, free, exact, egoless, and eternal. After three decades of research and practice, Alexander (2002–2005) eventually defined the order in some accurate mathematical language, and he used the term wholeness or living structure to capture the meaning of the kind of order (cf., Section 2 for more details). Living structure is defined as a mathematical structure, which consists of numerous recursively defined substructures with an inherent hierarchy. Across the hierarchy, there are far more small substructures than large ones, while on each level of the hierarchy the substructures are more or less similar in size (c.f., Appendix A for an example). The notion – or recurring notion – of far more smalls than larges has been formulated as the scaling law (Jiang 2015a), while the notion of more or less similar is often referred to as Tobler's law, or the first law of geography (Tobler 1970). These two laws are the fundamental laws of living structure or wholeness. In this paper, we treat geographic space as a living structure that consists of numerous recursively defined substructures for computing the degree of order.

In this paper, the term geographic space is used broadly, covering a wide range of scales between $10^{-2}$ (the size of a tiny ornament) and $10^6$ meters (the size of the Earth's surface). It can also refer to a map or an image of geographic space, the so-called table-top space (Egenhofer and Mark 1995). A space as a living structure, no matter how large or small, can always be decomposed into numerous substructures according to its internal structure and arrangement. The decomposition process ends up with far more smalls than larges across different hierarchical levels, but more or less similar on each of the hierarchical levels. In other words, a fine-structured whole may be seen as emerged and evolved, very much like organic things, from the rough-structured whole in a step-by-step fashion (see Figure 3 for illustration) to the fine-structured subwholes. Thus, a space is viewed as a hierarchy (or living structure) of nested subspaces, which are internally connected to constitute an organic whole. This conception of space differs fundamentally from current geographic representations in which geometric primitives (or mechanical pieces) such as points, lines, polygons, and pixels are only externally connected to form a mechanical totality (Goodchild et al. 2007) rather than an organic whole. The organismic way of representing space is substantially inspired by the philosophy of organisms (Whitehead 1929) and the third view of space formulated by Alexander (2002–2005): space is neither lifeless nor neutral, but a living structure capable of being more living or less living.

In this paper we develop a new approach to representing geographic space as a living structure, with which substructures are intrinsically (rather than externally) connected. The degree of order of a space is determined by three types of space: those spaces adjacent to it, the larger space that contains it, and those smaller spaces that are contained in it. In other words, a space is living (or orderly), if its adjacent spaces are living or orderly, the larger space to which it belongs is living or orderly, and what are contained in the space are also living or orderly. This degree of order that Alexander (2002–2005) conceived for characterizing living structure is essentially the same as the idea of PageRank (Page and Brin 1998), which is a variant of eigenvector centrality (Bonacich 1987) initially defined in social networks. The livingness or order of space is therefore a matter of fact rather than of opinion or personal preferences, so it provides an objective way of assessing goodness of space and its subspaces. In this connection, this paper can substantially contribute to sustainable urban planning and design.

The contribution of this paper can be seen from the following aspects: (1) a new approach to representing geographic space as a hierarchy of subspaces that are internally (rather than externally) connected to form a coherent whole, (2) a new kind of order of geographic space and a computational approach to the organic order, (3) a hierarchy of the 253 patterns and their degrees of order, and (4) computational evidence that traditional cities are more orderly than their modernist counterparts.

The remainder of this paper is organized as follows. In Section 2 we introduce related concepts, including the new kind of order and living structure, using very simple examples. Section 3 illustrates the new approach to representing geographic space as a hierarchy (or living structure) of recursively defined subspaces for computing the degree of order. To verify the approach, in Section 4 we report three case studies applied to the traditional city Venice and two modernist suburbs, 253 patterns, and 35 black-white strips. In Section 5 we further discuss the implications of the new geographic representation and the new kind of order on geographic information science (GIScience) and sustainable urban planning. Finally, Section 6 concludes and points to future work.

## 2. A new kind of order and living structure for characterizing geographic space

The new kind of order we deal with in this paper is a geometric order that exists pervasively in our surroundings, for example, in rooms, houses, gardens, streets, neighborhoods, and cities. The geometric order is so complex that no existing mathematics could be directly applied to it (Bohm 1969, 1980). In the three decades since the 1970s, there were two major developments in parallel. The first was led by the French mathematician Benoit Mandelbrot (1982) on fractal geometry, which is still very much framed under the Cartesian mechanistic worldview, while the other was led by the Austrian architect Christopher Alexander (2002–2005) on living geometry, under the organismic worldview (Whitehead 1920). Alexander conceived and developed the concept of order, a deep geometric order that people can use to create the kind of order in buildings, cities, paintings, and artifacts. The notion of order is tightly related to that of living structure, which is a mathematical structure that consists of numerous substructures with an inherent hierarchy: far more small substructures across the hierarchy (or the scaling law, Jiang 2015a), and more or less similar substructures on each of the hierarchy (or Tobler's law, Tobler 1970). In this section, we introduce the new kind of order and living structure (c.f., Appendix A for an example), as well as head/tail breaks (Jiang 2013a) and its induced index called the ht-index (Jiang and Yin 2014). Head/tail breaks is a recursive function for deriving the underlying living structure (or substructures) of a dataset, and the resulting substructures and their ht-index are used to quantify the degree of order.

To present the related concepts, let us use three simple datasets each with 10 numbers: [1, 1/2, 1/3, …, 1/10], [1+e1, 1/2+e2, 1/3+e3, …, 1/10+e10] (where ei denotes a very small value), and [1, 2, 3, …, 10]. The average value of the first dataset [1, 1/2, 1/3, …, 1/10] is around 0.29, which partitions the 10 numbers into the head for those greater than the average [1, 1/2, 1/3] and the tail for those less than the average [1/4, 1/5, …, 1/10]. For the three numbers in the head, their average is around 0.61, which further partitions the head [1, 1/2, 1/3] into the head [1], and the tail [1/2, 1/3]. Eventually, the notion of far more smalls than larges for the first dataset recurs twice, so the dataset is said to have a living structure with an inherent hierarchy, measured by the ht-index of three. This iterative partition process of the dataset into the head and the tail is therefore called head/tail breaks (Jiang 2013a). Generally speaking, head/tail breaks is a recursive function for iteratively deriving the notion of far more smalls than larges, until the notion is violated, e.g., the head percentage is over 40%.

The second dataset [1+e1, 1/2+e2, 1/3+e3, …, 1/10+e10] differs slightly from the first dataset with some very small values (ei), either positive or negative. The second dataset is said to follow Zipf's law (1949), a statistical regularity for characterizing city size distribution in a country or word frequency in a paper or book. For example, in a large enough country, the largest city is about twice as big as the second largest, approximately three times as big as the third largest, and so on. Like the scaling law, Zipf's law favors statistics over exactitude. Thus, the first dataset cannot be said to follow Zipf's law. For the second dataset, let us assume that the average is also around 0.29, which also partitions the 10 numbers into the head [1+e1, 1/2+e2, 1/3+e3] and the tail [1/4+e2, 1/5+e3, …, 1/10+e10]. We further assume that the second average is around 0.61, which again partitions the head [1+e1, 1/2+e2, 1/3+e3] into the head [1+e1] and the tail [1/2+e2, 1/3+e3]. The second dataset has the same hierarchy of three as the first, but the second, due to its randomness or the statistical nature, is more orderly or more living than the first. These two datasets provide a sense of feeling for the kind of order we deal with in the paper.

Unlike the first two datasets, which are nonlinear, the third dataset [1, 2, 3, …, 10] is linear. The average of the 10 numbers is 5.5, which partitions the dataset equally into two parts rather than the head and the tail. In other words, the third dataset lacks the notion of far more smalls than larges, not to mention the recurring notion of far more smalls than larges. Equivalently, the dataset is said to violate the scaling law (Jiang 2015a). Given this fact, the third dataset is therefore less orderly or less living than the first two. Also, unlike the first two datasets, the third dataset is said to have a nonliving or mechanical structure. To this point, we can rank these three datasets in terms of their degrees of order; the second dataset is the most orderly, followed by the first one and the third one. The reader may have sensed intuitively that the notion of order is tightly linked to that of living structure.

Having examined the three datasets in terms of their degrees of order, it is necessary to point out two kinds of order: organic and mechanical. Geographic curves such as coastlines belong to the first kind being organic or natural, while Euclidean geometric curves such as the straight line and the bowl-shaped curve the second being mechanical or artificial (Figure 1). Based on the concept of order illustrated above, geographic curves are more orderly than the Euclidean geometric ones. This is because with the geographic curves there is a recurring notion of far more small substructures than large ones. For example, for the curve shown in Figure 1b, the notion of far more small bends (or substructures in general) than large ones recurs twice, or with the ht-index being three. However, if the same curve is seen as a collection of individual segments, it tends to be deadly, for the individual segments are more or less similar in size. Conventionally, the Cartesian coordinate system provides a general mathematical framework for defining the mechanical order (Bohm 1969). Current geographic representations are essentially based on the Cartesian coordinate system, under which geographic features are usually viewed as deadly or with the mechanical order (c.f., Figure 2 for a further illustration). To a large extent, Alexander (2002–2005) has reversed our conventional ways of thinking on order. In this paper, we adopt the view that random curves are more orderly or more living than Euclidean geometric curves.

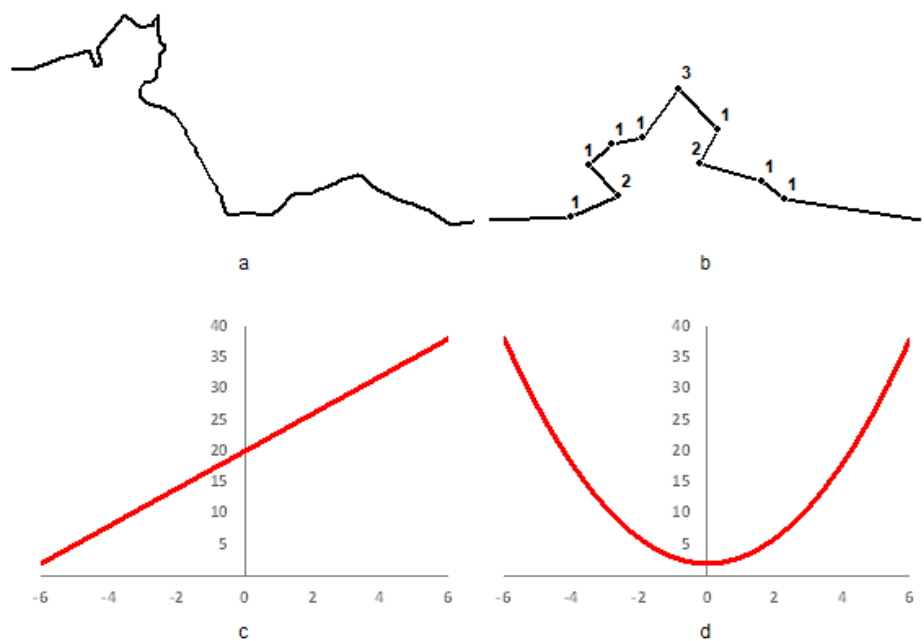

Figure 1: (Color online) Random curves are more orderly than mathematical curves
(Note: Random curves (a and b) contain far more small bends (or substructures) than large ones, so they are more orderly than the Euclidean geometric curves (c and d) that look pretty stiff. Instead of seeing the curve (b) as a set with 11 more or less similar sized line segments, it is seen as a set with far more small bends (or substructures in general) than large ones. The vertex 3 and the two ending points constitute the largest bend, followed by the second-largest bends marked by two, and by the third-largest bends marked by one.)

**3. A new geographic representation for computing the degree of order**
Any geographic space can be represented as a living structure with an inherent hierarchy: far more small substructures across the hierarchy, and more or less similar substructures on each level of the hierarchy. To illustrate this point, in this section we use the Dutch painter Piet Mondrian's famous painting composition to introduce the novel geographic representation, defined under the third view of space (Alexander 2002–2005). We present the two fundamental laws and the two design principles of living structure. We also clarify how the degree of order of subspaces can be computed using the PageRank algorithm (Page and Brin 1998), and the degree of order of the whole space is quantified by the substructures and their inherent hierarchy. That is, the more substructures the more orderly, and the higher hierarchy of the substructures the more orderly (Jiang 2019). Let us first look at how geographic space is represented in current geographic information systems (GIS).

Current geographic representations are based on object and field views of space, such as vector and raster (Figure 2), although many alternative representations have been developed for dealing with more complex and dynamic geographic phenomena (e.g., Goodchild et al. 2007, Yuan 2001). All of these current representations are essentially framed under the Cartesian mechanistic world view (Descartes 1637, 1954) or equivalently Newtonian absolute space and Leibnizian relational space. A geographic space is first mechanistically represented by a set of mechanical parts such as points, lines, polygons, and pixels, and then these mechanical parts are associated or externally connected. For example, the seven blocks in Figure 2b constitute adjacent relationships as indicated by the graph, and all of those pixels with the same colors and within the thick borders in Figure 2c are associated as same fields. At any moment, a geographic space is seen as static or lifeless under the current geographic representations. Instead of the static or lifeless view, we propose a dynamic and living view of space under the organismic world view.

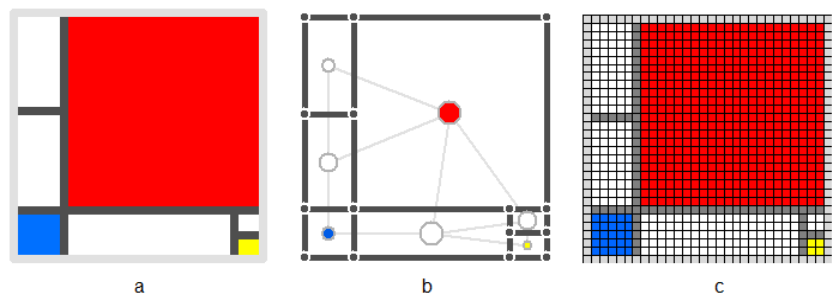

Figure 2: (Color online) Two conventional geographic representations based on object and field views (Note: A space (a) is represented mechanistically as vector (b) and raster (c) formats. The painting namely composition (a) was created by the Dutch painter Piet Mondrian (1872–1944), who is famous for his works of art using simple geometric shapes and primary colors of red, yellow, and blue. Two conventional geographic representations based on geometric primitives of points, lines, and polygons (b) and pixels (c).)

Table 1: Two fundamental laws of geography or living structure
(Note: The two laws complement rather than contradictory each other to characterize living structure or the Earth's surface.)

| Scaling law | Tobler's law |
|---|---|
| Far more small substructures than large ones across all scales | More or less similar substructures on each scale |
| Disproportionality between small and large (80/20) | Proportionality between small and large (50/50) |
| Without a characteristic scale | With a characteristic scale |
| Pareto distribution | Gauss-like distribution |
| Spatial heterogeneity or interdependence | Spatial homogeneity or dependence |
| Complex and non-equilibrium character | Simple and equilibrium character |

Under the organismic worldview, we see that the painting consists of 18 recursively defined substructures (Figure 3). In other words, the seven blocks of the painting are seen as emerged and evolved from the empty square in a step-by-step fashion, very much like cell division process or embryo-like growth (Panels a1–a4). The course of evolution is said to be governed by two fundamental laws of living structure: the scaling law and Tobler's law (Table 1). Clearly there are far more newborns (or newly generated substructures) than old ones from Panel a1 to a2, and again from Panel a2 to a3, although with a violation from Panel a3 to a4. On the other hand, from a design point of view, the empty square is differentiated in a step-by-step fashion into many substructures, which are well adapted to each other. Thus, there are two design principles of living structure: differentiation and adaptation, distilled from the 15 properties of living structure (Alexander 2002–2005). Note that a space is differentiated unequally rather than equally, so at each iteration there are far more small substructures than large ones. This is probably the key principle that the painter used. It is the same principle that makes the painting visually appealing or orderly. The notion of far more smalls than larges is more favored than that of more or less similar; The more differentiated, the more living or orderly, the higher hierarchy, the more living or orderly.

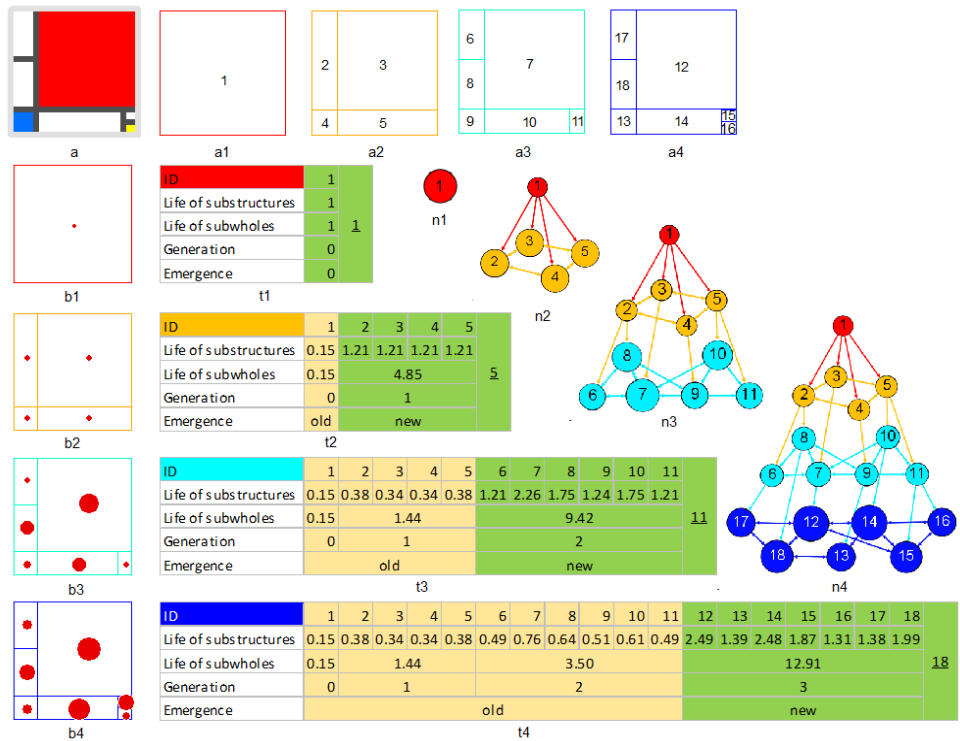

Figure 3: (Color online) The organic representation of the painting as a living structure
(Note: The painting, named *Composition*, was created by the Dutch painter Piet Mondrian (1872–1944), who is famous for using simple geometric shapes and primary colors (red, yellow, and blue). Mechanistically, the painting is composed of seven mechanical blocks, whereas organismically it consists of 18 recursively defined substructures, as shown in Panels (a1–a4). In this organismic way, the painting is viewed as emerged or evolved from the empty square (a1), in a step-by-step fashion indicated by generation 0 (red), 1 (orange), 2 (light blue), and 3 (blue). Interestingly, there are always far more newly generated substructures (in green or new) than old ones (in yellow or old) (t1–t4). All of the substructures constitute a topological network evolved from n1 to n4, where edges within each level of hierarchy indicate spatial proximity, while those across different levels of hierarchy represent emergence. From the topological network, the degree of order (or life) can be calculated (indicated by life in t1–t4) and visualized by the dots' size (b1–b4).)

The generated substructures constitute a coherent whole that is persistently transformed towards a living or more living structure – the painting itself – as shown in Figure 3. The transformation is illustrated in Panels a1–a4, and the corresponding degree of order (underlined numbers in Panels t1–t4) of the substructures is visualized in Panels b1–b4. Note that the degree of order is identical to the number of substructures, which means that the more substructures there are, the more orderly they are. This is the order of the whole, which increases from 1 to 5, 11 and finally to 18. The order score is distributed among the individual substructures, according to PageRank algorithm (Page and Brin 1998). Alexander (2002–2005) independently invented the method for computing the degree of order. His initial definition of the degree of order is as follows. A substructure is orderly – genuinely – if its adjacent substructures are orderly, the larger substructure to which it belongs is orderly, and what are contained in the substructure are also orderly. This recursive way of defining the degree of order is the same in spirit as that of ranking web pages: the important web pages are those to which many important web pages point (Page and Brin 1998). That is the reason why PageRank was adopted for computing the degree of order for individual substructures.

All substructures can be essentially placed into three categories: adjacent ones, larger one, and smaller ones. All the substructures constitute a topological network from which their degrees of order can be computed, and the results are shown in Panels t1–t4, and further visualized in Panels n1–n4 and b1–b4. To this point, we have computed the degree of order for both the living structure and its substructures. In addition to the score 18, the degree of order for the living structure can also be characterized by the ht-index (Jiang and Yin 2014, Jiang 2015b). That is, to take the order scores in t4 and apply the head/tail breaks 2.0 (with the 40% on average as the threshold) to them, the resulting ht-index is three. Both the order score (18) and the ht-index (3) accurately reflect the rule for measuring the degree of order: the more substructures, the more orderly, the higher hierarchy of the substructures, the more orderly.

Based on the number of substructures (S) and their inherent hierarchy (H), the degree of life (L) can be computed by the multiplication of S and H, i.e., $L = S \times H$ (Jiang and de Rijke 2021). For the scenario illustrated in Figure 3, the degree of order of the painting is $18 \times 3 = 54$ ($L1$). However, the painting has two more ways of decomposing into 20 substructures; instead of going from one empty square to the four substructures, it can be decomposed into first two substructures, and then to the four substructures. This way, two more substructures and one more hierarchical level are induced, so $L2 = L3 = 20 \times 4$. The degree of life (or order) on average for the painting is (54+80+80)/3 = 71.3.

## 4. Case studies

We conducted three case studies to verify the new approach and to demonstrate the advantages of the new representation for computing the degree of order. The first case study demonstrated that the traditional city Venice is more orderly than two modernist counterparts. The second case study aimed to show how the 253 patterns (Alexander et al. 1977) constitute a coherent whole and the smallest patterns possess the highest degrees of order, while the third case study ranked the 35 black-white strips that perfectly match the cognitive outcome. The key message of the case studies is that any space (no matter how big or how small it is, as small as a strip or as big as the planet) can be viewed as a coherent whole, and its degree of order can be well measured.

### 4.1 Three urban areas and their degrees of order

We examined three urban areas in Italy: central Venice and the two modernist suburbs of Rome Mirti and Morena. In terms of physical area, central Venice is only half the size of Mirti and about one-sixth the size of Morena (Table 2). However, in terms of the degree of order, the central Venice is three times more orderly than Morena and four times more orderly than Mirti (Table 2). In other words, the two modernist suburbs are far less living than the traditional counterpart. What makes the central Venice the most living among the three is the larger number of substructures (3,936) and their steep hierarchy (8). Looked at from the opposite side, what makes the modernist suburbs less living is the low number of substructures and their flat hierarchy (4 or 5) (Figure 4). There is little doubt that Venice is the most orderly among the three areas.

Table 2: The degree of order for the three urban areas (or cities)
(Note: The degree of order (L) is computed by the multiplication of the number of substructures (S), summed up from those at different levels of hierarchy (S0, S1, S2, … S7) and the hierarchical levels (H), indicating that Venice has the highest degree of order among the three cases. The hierarchy coloring is consistent with that in Figure 4.)

| Cities | Area (km²) | Levels of substructure | | | | | | | | Order | | |
|---|---|---|---|---|---|---|---|---|---|---|---|---|
| | | S0 | S1 | S2 | S3 | S4 | S5 | S6 | S7 | S | H | L |
| Venice | 0.71 | 1 | 2 | 4 | 9 | 28 | 89 | 989 | 2,814 | 3,936 | 8 | 31,488 |
| Mirti | 1.36 | 1 | 4 | 18 | 64 | 1,346 | | | | 1,433 | 5 | 7,165 |
| Morena | 4.01 | 1 | 4 | 43 | 2,462 | | | | | 2,510 | 4 | 10,040 |

According to the new representation shown in Figure 3, each of the urban areas is represented as a topological network consisting of all substructures at different levels of the hierarchy. By applying the PageRank algorithm, the degree of order for individual substructures can be computed. Table 3 shows overall results for those substructures at the lowest level of hierarchy. From the table, we can see that the most orderly substructures have the degree of order 4.7 for Venice, but they are 2.37 and 2.62 for Mirti and Morena, respectively. Venice is more densely packed than the two modernist areas; its area is only one-sixth that of Morena but has almost the same number of buildings. Interestingly, the three topological networks differ dramatically in terms of in-links (or degree of coming-in links), although the average degree of in-links is almost the same, around 6.83 or 6.85. However, the topological network of Venice is far more heterogeneous than those of the other two areas, and the degree of heterogeneity reflects different hierarchies among the three areas.

Table 3: The degree of order for the substructures at the lowest level of hierarchy
(Note: The degree of order is computed using the PageRank algorithm from the three topological networks for individual substructures.)

| Cities | | Area (km²) | Number | Area of substructure (m²) | | | Inlink | | | Order | | |
|---|---|---|---|---|---|---|---|---|---|---|---|---|
| | | | | Max | Average | Min | Max | Average | Min | Max | Average | Min |
| Venice | S7 | 0.71 | 2,814 | 9,643 | 252 | 3 | 28 | 6.85 | 2 | 4.7 | 1.21 | 0.35 |
| Mirti | S4 | 1.36 | 1,346 | 17,392 | 1,012 | 19 | 16 | 6.83 | 3 | 2.37 | 1.05 | 0.45 |
| Morena | S3 | 4.01 | 2,462 | 15,354 | 1,615 | 61 | 17 | 6.85 | 3 | 2.62 | 1.02 | 0.46 |

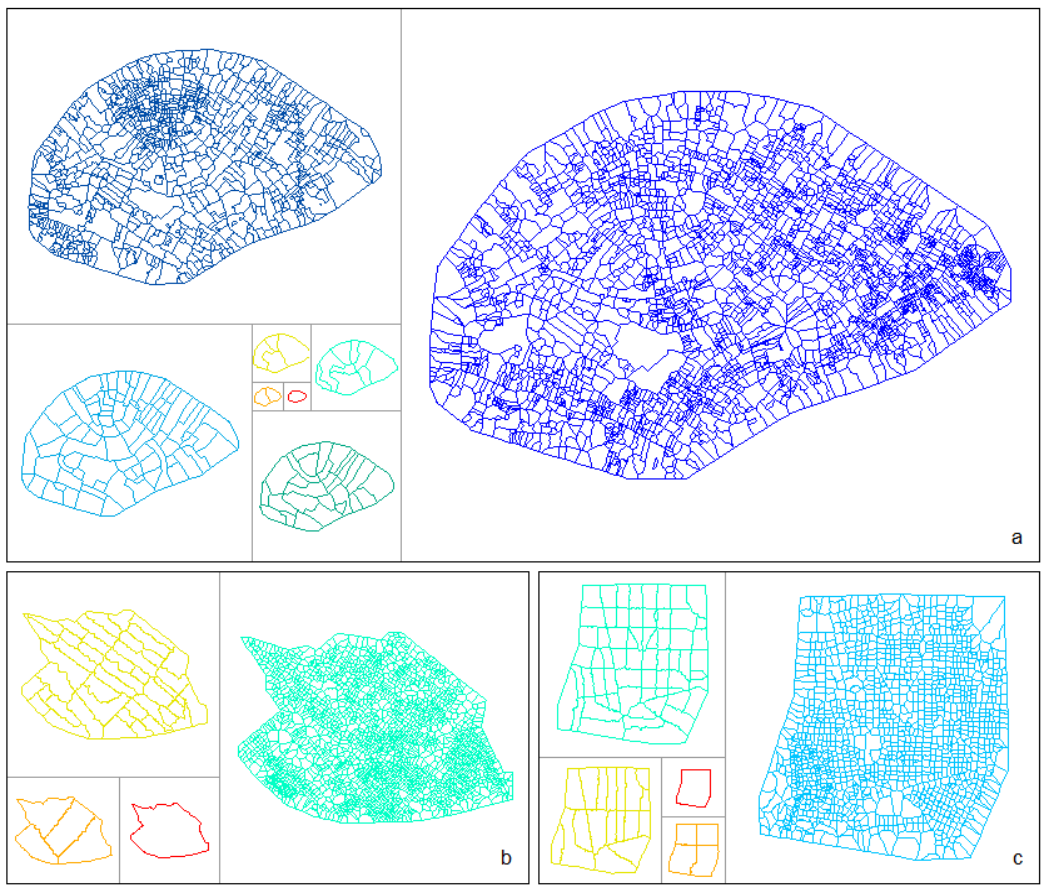

Figure 4: (Color online) Different levels of hierarchy for the three urban areas
(Note: A spiral-shaped layout is adopted for showing the different levels of hierarchy of the three urban areas (a) Venice, (b) Mirti, and (3) Morena, each of which is evolved from red towards blue. We focus on the underlying living structures or their configurations, so the actual sizes make little sense.)

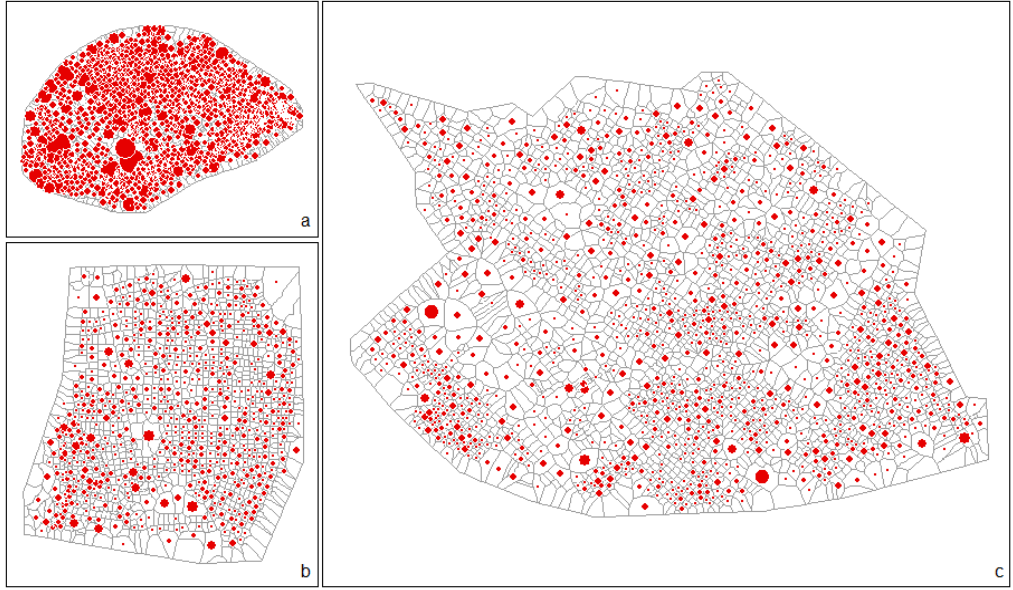

Figure 5: (Color online) Degree of order visualized by dot sizes
(Note: Unlike in Figure 4, three areas are proportional to their physical sizes, so we can see clearly that the area of Venice (a) is about a half that of Mirti (b), and one sixth of that of Morena (c).)

### 4.2 The 253 patterns as a coherent whole and their degrees of order

As mentioned earlier, the idea of living structure can be traced back to the classic pattern language work (Alexander et al. 1977), in which in total 253 patterns are defined and described. Each pattern describes a problem and a solution to the problem, and importantly how one pattern is connected to certain "larger" patterns, and to certain "smaller" patterns, just like the relationship of substructures. In this case study, we sought to apply the new representation to these patterns and to examine how the collection of more than 250 patterns form a coherent whole for computing their degrees of order. These patterns are ordered at different levels of scale (or hierarchy) ranging from the largest (world government, regions, and towns) to the smallest (things in our daily life) (Table 4). We identified the six hierarchical levels with the number of patterns increasing gradually from 1, for the largest world government, up to 145 for the smallest, clearly showing a scaling hierarchy (of living structure) with far more small patterns than large ones. In what follows, we describe in detail how the living structure was constructed and how the degree of order was then computed.

Table 4: The 253 patterns are organized into six hierarchical levels as a living structure
(Note: The patterns are treated as substructures of the living structure. The hierarchy coloring is consistent with that of Figure 6.)

| Hierarchy Coloring | Levels of scale (or hierarchy) Description | Number of people | Patterns Number | Order Sum | Order Average | Connectivity Inlink | Connectivity Outlink | Connectivity Sum | Connectivity Average |
|---|---|---|---|---|---|---|---|---|---|
| S0 | World government | All population | 1 | 0.18 | 0.18 | 0 | 7 | 7 | 7 |
| S1 | The region | 8,000,000 | 7 | 1.98 | 0.28 | 19 | 26 | 45 | 6.4 |
| S2 | The major city | 500,000 | 11 | 3.78 | 0.34 | 29 | 78 | 107 | 9.7 |
| S3 | Small towns, communities, and neighborhoods | 500 - 10,000 | 37 | 12.47 | 0.34 | 158 | 235 | 393 | 10.6 |
| S4 | House clusters and work communities | 30 - 50 | 52 | 23.78 | 0.46 | 311 | 499 | 810 | 15.6 |
| S5 | Families, room, person, ornament | 1 - 15 | 145 | 210.81 | 1.45 | 1273 | 945 | 2218 | 15.3 |

We first built up the topological network of these patterns according to their initial definition in the book in terms of which patterns are associated with other patterns. The first step is no different from previous works (Park 2015, Sousa et al. 2020). Unlike previous works, however, we then arranged the 253 patterns at the six hierarchical levels ranging from the largest of world government, regions, and towns to the smallest of families, room, person, and ornament (Table 4). Based on the hierarchical levels, we then edited some links to ensure that only two kinds of links are allowed: those at a same hierarchical level (links among brothers and sisters, so to speak) and those between two adjacent levels (links to father and sons, so to speak). In other words, links are not allowed between two hierarchical levels that are not adjacent. In this way, we slightly modified the network of the 253 nodes for computing their degrees of order.

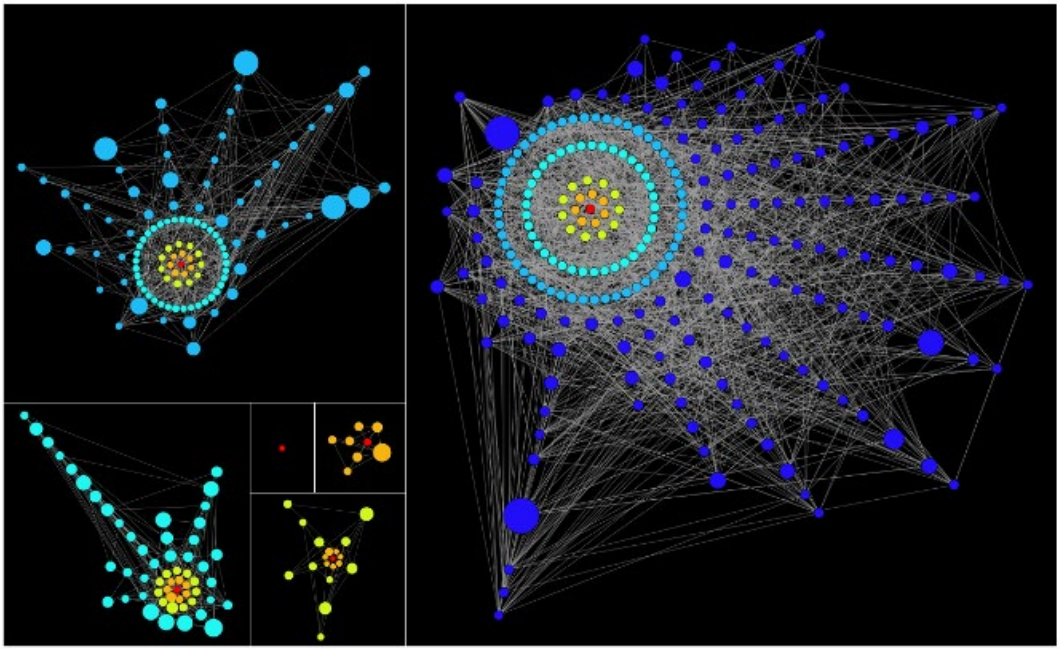

Figure 6: (Color online) Six hierarchical levels of the pattern language network
(Note: A spiral-shaped layout is adopted to show the inherent hierarchy of the 253 patterns. The coherent whole of the 253 patterns is considered to emerge and evolve from the single node of world government in a step-by-step fashion. Among the 253 patterns, the smallest ones tend to have the highest degree of order, indicating the design principle of differentiation.)

In this case study, patterns are considered to be equivalent to substructures, which together form a coherent or living structure. Figure 6 illustrates the network in its six hierarchical levels in a spiral-shaped evolution manner, beginning with the largest patterns to the smallest ones. The visualization in Figure 6 shows that the smallest patterns in the lowest hierarchical level tend to have highest degrees of order, reflecting the rule that the more differentiated, the more orderly. In addition, the degree of order is inherited from the earlier generation in a step-by-step fashion. Every new generation can be considered to be an emergence from the previous generation, so the degree of order is determined among the nodes of two consecutive generations. In each panel of Figure 6, we see that the next generation is more orderly than the previous one, and eventually the last generation has the highest degree of order. In other words, the next generation is an emergence of the previous generation.

The network of 253 nodes is an enhanced version of the network of 18 nodes for the painting composition, as illustrated in Figure 3. With the large network, we can note that the degree of order correlates with the recurring notion of far more smalls than larges. It is the recurring notion that makes Figure 6 visually appealing or beautiful – so-called structural beauty – which has little to do with the coloring. In other words, if we turn Figure 6 into a gray scale, the kind of beautiful structure remains.

To make a better sense of the case study, we may imagine in the future that our descendants may use the human wisdom of pattern language (Alexander et al. 1977) to start human settlements on Mars. They will begin individual settlements one by one, and as time goes, far more small settlements than large ones would be expected to emerge. At the building levels, there would be more and more human scales to emerge and develop, towards human daily life things. These human-made things must adapt to nature or naturally occurring things such as terrain surfaces, guided – either subconsciously or unconsciously – by the two design principles: differentiation and adaptation. This unfolding process of built environments on the surface of Mars is exactly what has happened on the Earth's surface over the past thousands of years in human civilization. Unfortunately, this natural process was interrupted over a century ago because of industrialization, and an inevitable outcome is the emergence of nonliving structures.

**4.3 The 35 strips and their degrees of order**
Let us examine how the new approach helps rank-order a set of the 35 strips with three black squares and four white ones (e.g., wwbbwwb) in terms of their degrees of order. In the following, we tried to avoid using squares, which focus only on the square perspective. Instead, we use bricks that could be either squares or rectangles. The strip experiment was initially conceived and carried out by Alexander and Carey (1968) to rank-order the set of strips in terms of their coherence and simplicity based on human perception tests. Interestingly, they found that the number of subsymmetries highly correlates with the human perception of coherence and simplicity. A subsymmetry is a symmetrical subsegment (Figure 7a). However, the way of accounting the number of subsymmetries is somewhat mechanistic, as it is based on individual squares, such as three subsymmetries with the length of two. This square perspective can be likened to the pixel perspective of an image. Instead of the mechanistic perspective, we adopted the organismic perspective as presented in the previous sections. Panels b and c of Figure 7 present two sets of substructures (which are brick-shaped) decomposed from the strip, in the same way as illustrated in Figure 3. Human perception is not based on the squares, but on the bricks or substructures. This is probably the major difference between substructures and subsymmetries. Therefore, the two adjacent white squares (or two adjacent black squares) are considered as a brick, being a whole brick or subwhole, so they should not be viewed as two symmetric squares, just like a set of adjacent pixels with the same color are only perceived as a homogeneous patch.

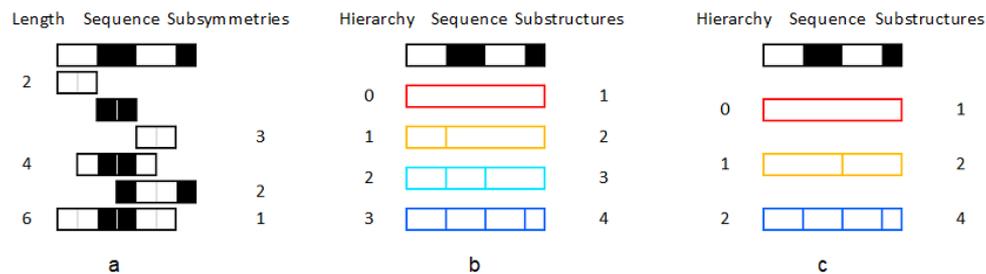

Figure 7: (Color online) Illustration of subsymmetries (a) and substructures (b and c)
(Note: The strip is decomposed into a unique set of six subsymmetries (a) and substructures (b and c). Panels b and c are just two possible ways of iteratively decomposing the strip into substructures until they are no longer decomposable, so called non-decomposable substructures.)

Let us examine in detail the notion of subsymmetries versus that of substructures. Alexander and Carey (1968) simply counted the number of subsymmetries as an indicator of the degree of order. For the strip in Figure 7, the total number of subsymmetries is $3 + 2 + 1 = 6$ (Figure 7a, which can be cross-checked in Figure 8). In contrast, we count not only the number of substructures (S), but also the hierarchy of the substructures (H), in order to compute the degree of order (or life); that is, $L = S \times H$ (Jiang and De Rijke 2021). According to the representations of Panels b and c, their degrees of order are, respectively, $L_b = (1 + 2 + 3 + 4) \times 4 = 40$, and $L_c = (1 + 2 + 4) \times 3 = 21$. In the end, the degree of order would be averaged from all possible decompositions. The resulting L is used to rank all 35 strips, as shown in Figure 8.

The ranking based on the computing result is very consistent with that of the cognitive experiments (Alexander 2002–2005, Appendix 3 of book 1); that is, beginning with those active (or living) patterns made of many small units to those lazy (or less living) ones with few long bars. The three groups – the left column, the right column, and the three in the middle – perfectly match those of the cognitive experiments. Now let us elaborate on the ranking and why it makes sense. First, the most orderly is number 15, while the least orderly is number 35 (or 1), because they contain seven and two pieces, respectively. To a large extent, the rank-order shown in Figure 8 can be interpreted by the number of bricks. It implies that the more complex, the more orderly as shown in Figure 1. In this regard, the notion of order is closely linked to the organized complexity (Jacobs 1961). Second, those strips that are mirrored to each other have the same degree of order; for example, the 35th is mirrored to the 1st (so their positions are interchangeable), as the underlying configuration or arrangement is the same. To a large extent, those mutually mirrored strips (e.g., the 35th and the 1st) have the same degree of order. This result helps verify the new representation and the new kind of order. We will further discuss their implications in the following section.

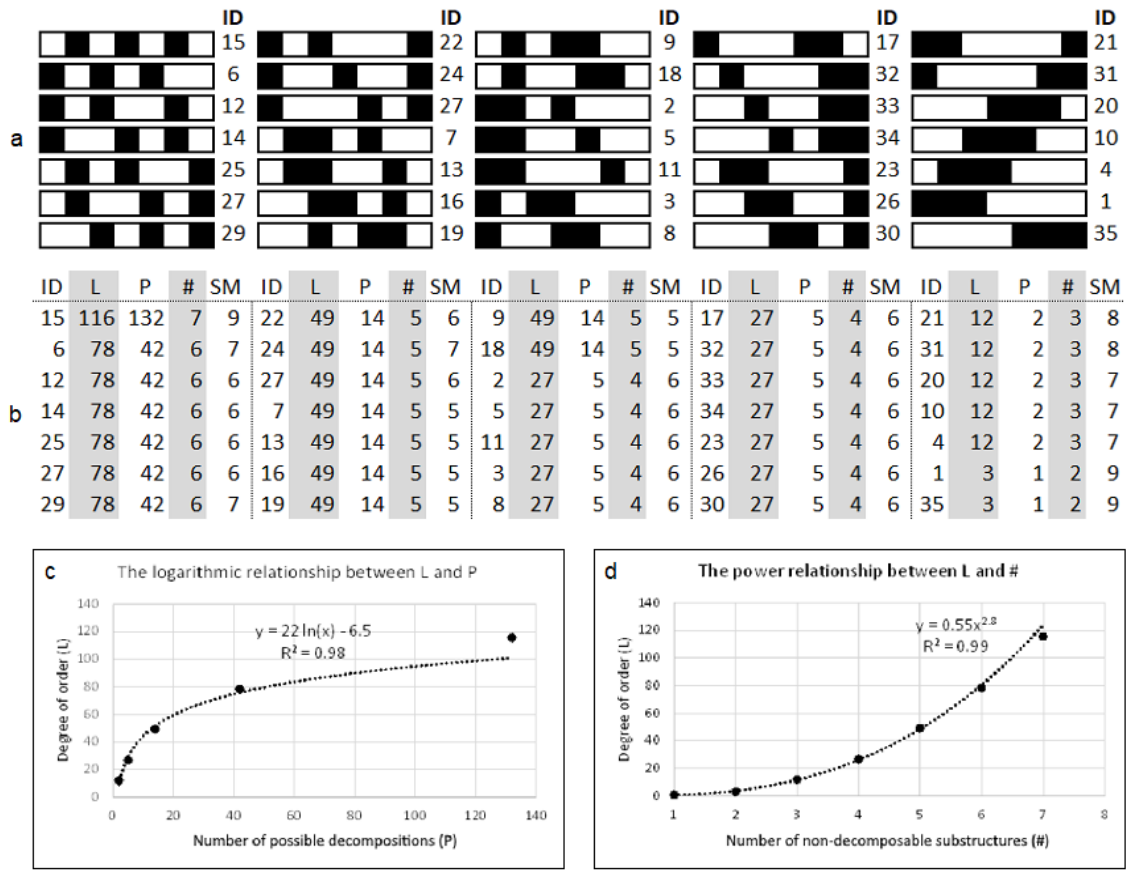

Figure 8: Ranking the 35 strips according to their degrees (L) of order
(Note: The 35 strips are ranked according to their degrees of order (L). This means the 15th strip is the most orderly, with the degree of order of 116, followed by those with the degree of order of 78, 49, 27, 12, and ended by the two strips with the degree of order of 3. This organismic way of ranking strips has been proved to be better than Alexander's initial mechanistic way of ranking based on subsymmetries (SM). P indicates all possible ways of iteratively decomposing a strip into substructures until they are no longer decomposable, while # represents the number of non-decomposable substructures. The degree of order (L) is stretched over a range between 3 and 116, which is much wider than that of for subsymmetries between 6 and 9.)

## 5. Implications of the new representation and the new kind of order

Under the organismic worldview, we put forward the new geographic representation to compute the degree of order inherent in geographic space. Compared to existing geographic representations (e.g., Yuan 2001, Goodchild et al. 2007, Jiang 2015b), in which spaces are externally rather than internally connected, the new representation that is built on the internal connection of recursively defined substructures shows several advantages. First, space bears an inherent hierarchy or spatial heterogeneity, so it can be hierarchically organized. Second, space or subspaces are organically represented as a living structure, which is in line with spatial cognition of human beings. In contrast, existing geographic representations tend to be based on "cold and dry" (paraphrased from Mandelbrot 1983) geometric primitives of points, lines, polygons, and pixels, which have little to do with what people perceive.

Third, the new representation enables us not only to understand city structure and dynamics, but also to transform cities – or geographic space in general – to be living or more living. Below we will further discuss the implications of this study on GIScience and sustainable urban planning. Before that, it is necessary to link the new representation or the hierarchy of substructures to some previous efforts.

Hierarchy is a key feature of various complex systems such as physical, biological, social, and informational. These complex systems are not made of pre-fabricated elements but are emerged and evolved towards coherent wholes or living structures. The complex systems or living structures are nearly decomposable into subsystems or substructures (Simon 1962). The notion of nearly decomposability underlies the decomposition process, being organic or natural, that we have demonstrated throughout the paper. It should be noted that the same notion of hierarchy has been used to develop data structure such as quadtree for geospatial database (Samet 2006), box-counting method for calculating fractal dimensions (Mandelbrot 1983), and the subdivision method for showing different levels of hierarchy of cities (e.g., Chen and Wang 2014). Along this line of research, the decomposition process is mechanical or unnatural rather than organic or natural.

The new representation is an organic representation in the sense that it organically or naturally reflects how human beings perceive or understand geographic space. Space at any moment is not lifeless but a living structure capable of being more living or less living. It can help deal with a series of problems in GIScience, such as map generalization, visualization, data classification, and spatial cognition. Map generalization can be simply viewed as the reverse of the unfolding process from the top hierarchy to the low hierarchy. For example, the painting composition is viewed as the unfolding process from 1 to 4, from 4 to 6 and from 6 to 7, while the reverse process from 7 to 6, from 6 to 4, and from 4 to 1 is generalization. It is essentially the inherent hierarchy of geographic space that governs map generalization and classification. Spatial cognition or the image of the city (Lynch 1960, Jiang 2013b) is largely determined by the underlying living structure. In other words, those substructures with the highest degree of order or the most salient substructures tend to constitute the image of the city.

Sustainable urban planning is not a LEGO-like assembly of prefabricated elements, but an embryo-like growth with persistent adaptation towards a coherent whole or living structure (Thompson 1917, Alexander 1987, Alexander 2002–2005, Jiang and Huang 2021). Given the adaptive nature of city growth, cities are essentially unpredictable (Alexander 1987, Batty 2018), but they ought to become living or more living. In this connection, the new representation provides an effective means to assess the livingness of cities or the degree of order of cities. Unfortunately, the new kind of order has not been well recognized in the literature of urban planning. Instead, the kind of order is often dismissed as disorder or without order, for it looks irregular on the surface. This kind of order can also be called hidden order, which is able to trigger a sense of livingness or beauty in the human mind and heart. Unfortunately, the mainstream urban planning is still misguided by mechanical order or *"cold and dry"* geometric shapes rather than living structures.

We have demonstrated that a higher degree of order lies in details or small substructures rather than large ones. The small or smaller substructures come from persistent differentiation. This is a revelation for sustainable urban planning, *not* to expand a city (to be sprawl), but to make it more differentiated, with far more smalls than larges. This result provides evidence of why geometrical fundamentalism (Mehaffy and Salingaros 2006) is so wrong, for it focuses on Plato geometrical shapes (either too large or too fragmented) that often violate the scaling law (or the recurring notion of far more small substructures than large ones) and is based on mechanistic or assembly view rather than the organic growth view, or they do not come from differentiation. This work may significantly contribute to geodesign or the science of design in general (Fisher et al. 2020, Simon 1988) from the unique structural perspective, for it provides an objective way of quantifying the degree of order or the degree of goodness. In other words, goodness of design is no longer considered to be an opinion or personal preferences, but a matter of fact.

## 6. Conclusion

There are two different views of seeing the world: the Cartesian mechanistic worldview, under which things are seen as a collection of mechanical pieces, and the Whiteheadian organismic worldview, which sees things towards a coherent whole or living structure. Under the organismic worldview, space is seen as neither lifeless nor neutral, but as a living structure capable of being more living or less living; this is the so-called third view of space (Alexander 2002–2005). The present paper has developed a new approach to representing geographic space as a hierarchy of recursively defined subspaces for computing the degree of order, being organic rather than mechanical. These defined subspaces are internally rather than externally connected to form a coherent whole with a certain degree of order. We conducted three case studies to verify the computational approach and demonstrated three findings: (1) the city Venice is more orderly than the two modernist suburbs; (2) the smallest patterns have the highest degree of order, implying that any space should be further differentiated (with far more smalls than larges) to reach a higher degree of order; and (3) the ranking of the 35 strips based on their degrees of order perfectly matches the cognitive result. This new approach is superior to existing ones in the following ways: (1) it can show explicitly the inherent hierarchy or spatial heterogeneity, (2) it is consistent with spatial cognition of human beings, and (3) it helps understand not only how cities are, but also what cities ought to be. Therefore, it helps to effectively deal with many long-standing problems, such as map generalization and sustainable urban planning or design.

The new kind of order defined in this paper is not mechanical or Euclidean geometric order, but organic or natural order, which may look chaotic or random on the surface. The kind of order lies on the underlying living structure and exists not only in nature (naturally occurring things), but also in what we human beings make and build such as artifacts, buildings, and cities. The kind of order is a physical phenomenon that can be mathematically defined and can well be reflected in the human mind and heart, triggering a sense of order. The order can also be called livingness, beauty, coherence, life, or harmony. The new kind of order, with help of the new representation, can be well computed for guiding sustainable urban planning through the two design principles of differentiation and adaptation. Urban environments must be further differentiated to create more substructures, and the created substructures must be well adapted to each other, not only locally but also globally. In this connection, the 15 properties under the theory of living structure (Alexander 2002–2005) can serve as the transformation properties to make cities and communities living or more living towards a sustainable society. Our future work points to neuroscientific evidence on how the kind of order is perceived or reflected in the human mind and heart.

**Data and code availability statement**
The data used and generated in this study are packed and put publicly accessible at the Figshare site: https://figshare.com/s/2eec08052a96bd806772. The package includes (1) Python scripts for recursively decomposing a black and white strip into substructures until they are no longer decomposable, (2) Excel files with results of the Mondrian painting analysis, and of the strip decomposition processes, and (3) Shapefiles for the decomposition results of the three urban areas Venice, Mirti and Moren). The scripts are based on python 2.7 (https://www.python.org/download/releases/2.7/). The software tools used in the study are ArcGIS 10.8 (https://www.esri.com/en-us/arcgis/products/arcgis-desktop/overview) for the processing and analysis of spatial data, Microsoft Excel (https://www.microsoft.com/en-us/microsoft-365/excel), and Gephi 0.9.2 for network analysis (https://gephi.org/).

**Appendix A: Glossary of terms used in the paper**

We put together in this Appendix several terms that are frequently referred in the paper to make the paper more accessible. For the sake of simplicity, we use a tree – either dead or alive – to explain and exemplify the terms including hierarchy, order, life, coherence, structural beauty, living structure, substructures, symmetry, subsymmetries, complexity, and complex systems. The Appendix is intended to be self-contained, although we still point to some sources if appropriate.

Imagine a tree that consists of many individual subsets at the different levels of hierarchy, i.e., one trunk, two limbs, eight branches, 24 twigs, and 96 leaves. Apparently, the tree is with an inherent hierarchy of far more smalls than larges. To be specific, there are five hierarchical levels, namely trunk, limbs, branches, twigs, and leaves, so the tree is a living structure, no matter it is dead or alive biologically. Formally, living structure is defined as a structure in which there is a recurring notion of far more small substructures than large ones. The substructures are to the living structure what the subsets are to the tree. The tree can be viewed to emerge and evolve from the trunk to the limbs, from the limbs to the branches, from the branches to the twigs, and from the twigs to the leaves in a step by step fashion. In each step of the evolution, there is a notion of far more newborns than old ones. Because of the inherent hierarchy or living structure, the tree is said to be orderly or with a high degree of order (or life or coherence or structural beauty). It is commonly conceived that beauty is in the eye of the beholder. However, Alexander (2002–2005) in his life's work *The Nature of Order* challenged the conventional wisdom and argued that beauty is largely structural, so there is a shared notion of beauty among people and even different peoples regardless their cultures, gender, and races. Most people think the tree is beautiful. The concept of structural beauty is intended to capture the underlying living structure of space or things in general.

The tree is a complex system, so it is with a high degree of complexity, or organized complexity to be more precise (Jacobs 1961). Many complex systems such as physical, biological, social, and informational are with an inherent hierarchy (Simon 1962). The degree of complexity of the tree can be measured by its elements or subsets ($1 + 2 + 8 + 24 + 96 = 134$), and by its hierarchical levels (5). The complexity is therefore calculated by the multiplication of $134 \times 5 = 670$, so complexity is synonymous to order, life, coherence, or structural beauty. The readers who are familiar with fractals may refer to the tree as a fractal (Mandelbrot 1982). It is indeed true! However, we in this paper refer to the tree as a living structure, for it is defined under the Whiteheadian organismic worldview, while fractal may be defined under the Cartesian mechanistic worldview.

Subsymmetries are subsets of a symmetry, while substructures are subsets of a living structure. The tree is said to be with a symmetrical structure, symmetrical globally. It is important to note that subsymmetries or local symmetries are more important than global symmetry, for they determine the degree of global symmetry. This notion of symmetry with a focus on subsymmetries is very much like that of living structure from the recursive point of view. That is, if there are many subsymmetries, then the global structure tends to be very symmetrical; if there are many substructures, then the overall structure tends to be very living or orderly (Alexander and Carey 1968). This notion of symmetry with a focus on local symmetries is deeper than that of global symmetry such as translational, rotational, reflexive and glide (Weyl 1952, Wade 2006). It should be noted that we use the tree as a simple example of living structure, but we are not advocating the idea that a space should be designed to be with a tree structure, not at all. As famously claimed by Alexander (1965) in his classic work, a city cannot, should not, and must not be designed as a tree. Instead, a city or space in general should be designed as a complex network (Newman et al. 2006), as demonstrated in this paper.